\documentclass{emulateapj}

\newcommand{\axpa}{\hbox{CXOU\,J010043.1-721134}}
\newcommand{\axpb}{\hbox{4U\,0142+61}}
\newcommand{\axpc}{\hbox{1E\,1048.1-5937}}
\newcommand{\axpd}{\hbox{1RXS\,J170849-400910}}
\newcommand{\axpe}{\hbox{XTE\,J1810-197}}
\newcommand{\axpf}{\hbox{1E\,1841-045}}

\newcommand{\axph}{\hbox{1E\,2259+586}}
\newcommand{\sgra}{\hbox{SGR\,0526-66}}

\newcommand{\sgrc}{\hbox{SGR\,1806-20}}
\newcommand{\sgrd}{\hbox{SGR\,1900+14}}
\newcommand{\rxb}{\hbox{RX\,J0720.4$-$3125}}

\newcommand{\rbsd}{\hbox{RBS1223}}

\usepackage{natbib}
\shorttitle{The impact of magnetic field on NS cooling}
\shortauthors{Aguilera, Pons, \& Miralles}

\begin{document}
\title{The impact of magnetic field on the thermal evolution of neutron stars}
\author{Deborah~N.~Aguilera\altaffilmark{1,2},
Jos\'e A. Pons\altaffilmark{1},
Juan A. Miralles\altaffilmark{1}}
\altaffiltext{1}{Departament de F\'{\i}sica Aplicada, Universitat d'Alacant, 
             Apartat de Correus 99, E-03080 Alicante, Spain}
\altaffiltext{2}{Theoretical Physics, 
Tandar Laboratory, National Council on Atomic Energy, (CNEA-CONICET), Av. Gral. Paz 1499, 
1650 San Mart\'in, Pcia. Buenos Aires, Argentina}
\begin{abstract}
The impact of strong magnetic fields $B>10^{13}$G on the thermal evolution
of neutron stars is investigated, including crustal heating by magnetic field
decay. For this purpose, we perform
2D cooling simulations with anisotropic thermal conductivity
considering all relevant neutrino emission processes for realistic neutron stars.
The standard cooling models of neutron stars are called into question by showing that the magnetic
field has relevant (and in many cases dominant) effects on the thermal evolution. 
The presence of the magnetic field significantly  affects the thermal surface distribution and the
cooling history of these objects 
during both, the early neutrino cooling era and the late photon cooling era.
The minimal cooling scenario is thus more complex than generally assumed. A consistent
magneto-thermal evolution of magnetized neutron stars is needed to explain
the observations.

\end{abstract}


\maketitle
 

It has been long hoped that the comparison of theoretical models
for the cooling of neutron stars (NSs) with the direct observation of their thermal emission 
would help to unveil the physical conditions in the interior of these fascinating objects
\citep{Page2004,Yakovlev2004}.
Our knowledge of the cooling history of a NS has been improving as we 
were refining the physical ingredients that play a
key role on the thermal evolution of NSs. 
In the past, the field has become more
exciting every time that a new relevant idea was introduced 
(direct Urca, superfluidity in dense matter,
fast processes due to exotic matter, etc.). However, despite the fact that a number of
NSs are known to have large magnetic fields, most studies assumed 
weak magnetic fields. The main reason for
this simplification was that
the observed distribution of magnetic fields in radio-pulsars peaks in a region where
its effect were thought not to be relevant.

The increasing evidence that most of the nearby NSs with reported thermal emission 
in the x-ray band of the electromagnetic spectrum have anisotropic 
surface temperature distributions \citep{Zavlin2007,Haberl2007},
the striking appearance of {\it magnetars} \citep{Kaspi2007},
and the discovery of thermal emission
from some high field radio-pulsars \citep{Gonzalez2005}, are indicating that most
NSs which can be potentially used to contrast theoretical
cooling curves have actually large magnetic fields  ($B>10^{13}$ G).
The conclusion is that a realistic NS cooling model must not avoid
the inclusion of high magnetic fields.

The so-called {\it minimal cooling} scenario
\citep[see e.g.][for recent reviews]{Page2004,Yakovlev2004}
defines the cooling model in which the emissivity is given by
slow processes in the core, such as modified Urca and nucleon--nucleon Bremsstrahlung,
and enhanced by the neutrino emission from the formation and breaking
of Cooper-pairs of superfluid neutrons.
On the other hand, if fast neutrino processes (i.e. direct URCA) take place, 
the evolution of a NS changes
dramatically, resulting in the {\it enhanced} or {\it fast cooling} scenario.
Nevertheless, direct URCA only operates in the inner core of high mass NSs 
for some equations of state.

In this letter, we want to revisit the {\it minimal cooling} 
model considering the effects of magnetic field. If a {\it minimal model} must include
the minimum number of ingredients (but all the necessary ones) to explain the observations, the magnetic field should be taken into account as well.

The effect of the magnetic field on the surface temperature distribution caused by the 
anisotropic heat transport in the envelope was studied in 
a pioneering paper by \citet{Greenstein1983}.
The observational consequences of these models were analyzed for the pulsars 
Vela and Geminga among others \citep{Page1995}. 
 \cite{Potekhin2001} calculated the angular distribution of temperatures
in magnetized envelopes
taking into account the quantizing effect of the magnetic field on the electrons,
and the suppression (enhancement) of the electron thermal conductivity
in the direction perpendicular (parallel) to magnetic field lines. 
Nevertheless, the anisotropy generated in the envelope is not strong enough to be 
consistent with the observed thermal distribution of some isolated NSs,
and it should be originated deeper in the NS crust.
The understanding of kinetic properties of matter
in NS crusts and envelopes has also been recently improved, 
with special attention received by
the role of ions and phonons \citep{CH2007}, which can be relevant at low
temperatures and densities. In addition, the effect of impurities 
on the heat conduction in a non--perfect
lattice is also an open problem that must be considered in the near future.

More recently, crustal confined magnetic fields were considered to be responsible for the 
surface thermal anisotropy observed in some isolated NSs.
Temperature distributions in the crust were obtained as 
stationary solutions of the diffusion equation with axial symmetry \citep{Geppert2004}. 
The approach assumed an isothermal core 
and a magnetized envelope as inner and outer boundary condition, respectively. 
The results showed important deviations from the crust isothermal case for crustal confined magnetic 
fields with strengths $B>10^{13}$ G and temperature $T< 10^{8}$ K. 
Same conclusions have been obtained
considering not only poloidal but also toroidal components of the magnetic field 
\citep{Azorin2006, Geppert2006}.  These models succeeded in explaining simultaneously 
the observed x-ray spectrum, the optical excess, the pulsed fraction, and other 
spectral features of some isolated NS, such as RX J0720.4-3125 \citep{Perez2006}. 

Although former studies about anisotropic temperature distributions 
on the cooling history of NSs \citep{ShibYak1996,Potekhin2001} 
provided very useful information, detailed 
investigations of heat transport in the non spherical case for different
magnetic field geometries have not been available until recently.
In \cite{Aguilera2007} we have revisited the cooling of NSs combining the
insulating effect of strong non radial fields with the additional source of heating 
due the Ohmic dissipation of the magnetic field in the crustal region.
We have shown that, during the neutrino cooling era and the early stages of the 
photon cooling era, the thermal evolution is coupled to the magnetic field evolution, 
and both processes (cooling and magnetic
field diffusion) proceed on a similar timescale ($\simeq 10^{6} $ yr). 
The energy released by magnetic field decay (Joule heating) in the crust
is an important heat source that modifies or even controls the thermal evolution of a NS.
There is indeed observational evidence of this fact.
As shown in \cite{PonsLink2007}, there is a strong correlation between the inferred 
magnetic field and the surface temperature in a wide range of magnetic fields: 
from magnetars ($B \geq 10^{14}$ G), through radio-quiet isolated NSs
($B \simeq 10^{13}$ G) 
down to some ordinary pulsars ($B \leq 10^{13}$ G). The main conclusion is that,
rather independently from the stellar structure and the matter composition, 
the correlation can be explained by heating from dissipation of currents
in the crust on a timescale of $\simeq 10^{6}$ yrs.
\begin{deluxetable}{lcccr}
\tablecolumns{5} 
\tablewidth{0pc} 
\tablecaption{Properties of neutron stars with reported thermal emission, ordered
by decreasing magnetic field strength.}
\tablehead{
 \colhead{Source}& \colhead{$T^{\infty}$}  & \colhead{$t_{sd}$}    & \colhead{$B_{d}$} & \colhead{Ref.}  \\
 \colhead{ }     & \colhead{($10^6$ K)}  & \colhead{($10^3$ yr)} & \colhead{(10$^{13}$ G)} & \colhead{} 
}
\startdata
\sgrc &$7.56^{+1.6}_{-0.7}$& 0.22& 210& 1 \\\noalign{\smallskip}
\sgra &$6.16^{+0.07}_{-0.07}$& 2.0 & 73 & 1 \\\noalign{\smallskip}
\axpf &$5.14^{+0.02}_{-0.02}$& 4.5 & 71 & 1 \\\noalign{\smallskip}
\sgrd &$5.06^{+0.93}_{-0.06}$& 1.1 & 64 & 1 \\\noalign{\smallskip}
\axpd &$5.30^{+0.98}_{-1.23}$& 9.0 & 47 & 1 \\\noalign{\smallskip}
\axpc &$7.24^{+0.13}_{-0.07}$& 3.8 & 42 & 1 \\\noalign{\smallskip}
\axpa &$4.44^{+0.02}_{-0.02}$       & 6.8 & 39 & 1 \\\noalign{\smallskip}
\axpe &$7.92^{+0.22}_{-5.83}$& 17  & 17 & 1 \\\noalign{\smallskip}
\axpb &$4.59^{+0.92}_{-0.40}$& 70  & 13 & 1 \\
\hline\noalign{\smallskip}
PSR J1718-3718         &$1.69^{+0.62}_{-0.23}$&34   &7.4 &2 \\\noalign{\smallskip}
\axph                  &$4.78^{+0.34}_{-0.89}$&230  &5.9 &1 \\\noalign{\smallskip}
CXOU J1819-1458        &$1.40^{+0.47}_{-0.47}$&117  &5.0 &3 \\\noalign{\smallskip}
PSR J1119-6127         &$2.40^{+0.30}_{-0.20}$&1.7  &4.1 &4 \\\noalign{\smallskip}
\rbsd                  &$1.00^{+0.0}_{-0.0}$&1461 &3.4 &5 \\\noalign{\smallskip}
\rxb                   &$1.05^{+0.06}_{-0.06}$&1905 &2.4 &5 \\
\hline\noalign{\smallskip}
PSR B2334+61           &$0.65^{+0.13}_{-0.34}$&40.9 &0.99 &6 \\\noalign{\smallskip}
PSR B0656+14           &$1.25^{+0.03}_{-0.03}$&111  &0.47 &7 \\\noalign{\smallskip}
PSR B0531+21 (Crab)    &$<1.97$               &1.24 &0.38 &8  \\\noalign{\smallskip}
PSR J0205+6449         &$<1.02$               &5.37 &0.36 &9  \\\noalign{\smallskip}
RX J0822-4300          &$1.75^{+0.15}_{-0.15}$&7.96 &0.34 &10 \\\noalign{\smallskip}
PSR B0833-45 (Vela)    &$0.68^{+0.03}_{-0.03}$&11.3 &0.34 &11 \\\noalign{\smallskip}
PSR B1706-44           &$0.82^{+0.01}_{-0.34}$&17.5 &0.31 &12 \\\noalign{\smallskip}
PSR J0633+1748         &$0.50^{+0.01}_{-0.01}$&342  &0.16 &7  \\\noalign{\smallskip}
PSR B1055-52           &$0.79^{+0.03}_{-0.03}$&535  &0.11 &7  \\
\enddata
\tablerefs{(1) SGR/AXP Online 
Catalogue\footnote{http://www.physics.mcgill.ca/~pulsar/magnetar/main.html}; 
(2) \cite{Kaspi2005}; (3) \cite{Reynolds2006}, \cite{McLaughlin2007}; 
(4) \cite{Gonzalez2005}; (5) \cite{Haberl2007}; (6) \cite{McGowan2006}; 
(7) \cite{DeLuca2005}; (8) \cite{Weisskopf2004}; (9) \cite{Slane2004};
(10)\cite{Zavlin1999}, \cite{Hui2006}; (11) \cite{Pavlov2001}; 
(12) \cite{McGowan2004}. 
For the PSRs, $t_{sd}$ and $B_{d}$ from Pulsar Online 
Catalogue\footnote{http://www.atnf.csiro.au/research/pulsar/psrcat/}. 
}
\end{deluxetable}

In order to investigate if there is observational evidence that supports our models for
the crustal field evolution, we have compared theoretical cooling curves including magnetic 
field decay with a sample of NSs with reported thermal emission. To obtain the cooling curves 
we have performed two--dimensional simulations by solving
the energy balance equation that describes the thermal evolution of a NS 
\begin{equation}
C_{v}  \frac{\partial T}{\partial t}  - \vec{\nabla} \cdot
( \hat{\kappa} \cdot \vec{\nabla}  T) = - Q_{\nu} + Q_{\rm J}~,
\label{eneq}
\end{equation}
where $C_v$ is the specific heat per unit volume, $Q_{\nu}$ are energy losses
by $\nu$-emission, $Q_{\rm J}$ the energy gains by Joule heating, 
and $\hat \kappa$ is the thermal conductivity tensor, in general anisotropic in presence of
a magnetic field. In this equation we have omitted relativistic factors for simplicity.
A detailed description of the formalism, the code, and results can be found in
\cite{Aguilera2007}. The geometry of the magnetic field is fixed during the evolution.
As a phenomenological description of the field decay, we have assumed the following law 
\begin{equation}
B= B_0\frac{\exp{(-t/\tau_{\rm Ohm})}}
{1+(\tau_{\rm Ohm}/\tau_{\rm Hall})
(1-\exp{(-t/\tau_{\rm Ohm})})}~,
\end{equation}
where $B$ is the magnetic field at the pole, $B_0$ its initial value,
 $\tau_{\rm Ohm}$ is the Ohmic characteristic time,
and $\tau_{\rm Hall}$ the typical timescale of the fast, initial Hall stage.
In the early evolution, when $t\ll \tau_{\rm Ohm}$,
we have $B \simeq  B_0 (1+t/\tau_{{\rm Hall}})^{-1}$
while for late stages, when $t \geq \tau_{\rm Ohm}$,
$B \simeq  B_0 \exp(-t/\tau_{{\rm Ohm}})$.
This simple law reproduces qualitatively the results from more complex simulations
\citep{PonsGeppert2007} and facilitates the implementation of field decay
in the cooling of NSs for different Ohmic and Hall timescales.

Our sample of objects includes magnetars, isolated radio-quiet NSs, 
and radio-pulsars, and it covers
about three orders of magnitude in magnetic field strength. 
Before we discuss our results several caveats should be mentioned.
For most NSs, $B$ is estimated by assuming that the lose of angular momentum
is entirely due to dipolar radiation. If the magnetic field is constant
along the evolution, the dipolar component can be estimated by
$B_d = 3.2 \times 10^{19} ({P \dot{P}})^{1/2}$ G, where $P$ is the
spin period in seconds, and $\dot{P}$ is its time derivative.
Alternatively, for a few radio-quiet isolated NSs,
$B$ is estimated assuming observed x-ray absorption features are proton cyclotron
lines. This association is still controversial and, if true, it 
should be read as an estimate of the surface field, which is usually larger than the 
external dipolar component. In order to work with a more homogeneous sample,
we have included in our study only those objects for which
$\dot{P}$  is available. The quoted magnetic fields are always the spin-down
estimate for the dipolar component.

Ages are also subject to a large uncertainty.  
If the birth spin rate far exceeds the present spin rate and $B$ is considered constant, 
the age can be estimated by the {\it spin-down age} ($t_{sd}=P/2\dot{P}$). 
In the cases that an independent estimate is available (e.g. a kinematic age), the spin-down age 
does not necessarily coincide with the other estimation.
If one considers magnetic field decay, however, $t_{sd}$ seriously
overestimates the {\it true age} ($t$).
Correcting the spin-down age to include the temporal variation of the magnetic field
may help to reconcile the observed discrepancy between
the spin-down ages and independent measures of the ages of some of the objects we have 
included in this analysis.
This makes the comparison with observations more meaningful.
Other sources of error do probably exist but it is not
the purpose of this letter to discuss each observation separately.

As for the temperatures, in general there is no clear way to define error-bars. We have adopted
the range of values found in the literature, and we also indicate those cases in which the
estimate should be interpreted as an upper limit, more than a measure.
This is the case of the Crab pulsar, or even of some magnetars, which show 
large variations in the flux in the soft x-ray
band on a timescale of a few years, indicating that the {\it real} thermal component may be
that measured during quiescence and that the x-ray luminosity during their active periods 
might be a result of magnetospheric activity. Following the criteria adopted
in \cite{Page2004}, most reported temperatures are blackbody temperatures, except for 
low field radio-pulsars in which the blackbody fit results in an unrealistic
small radius of the NS. In these four cases, we took the temperature consistent with
Hydrogen atmospheres.
We include a list of the sources considered in Table 1, with the corresponding references.
\begin{figure}[thb]
\centering
\vspace{0.8cm}
\includegraphics[width=0.45\textwidth]{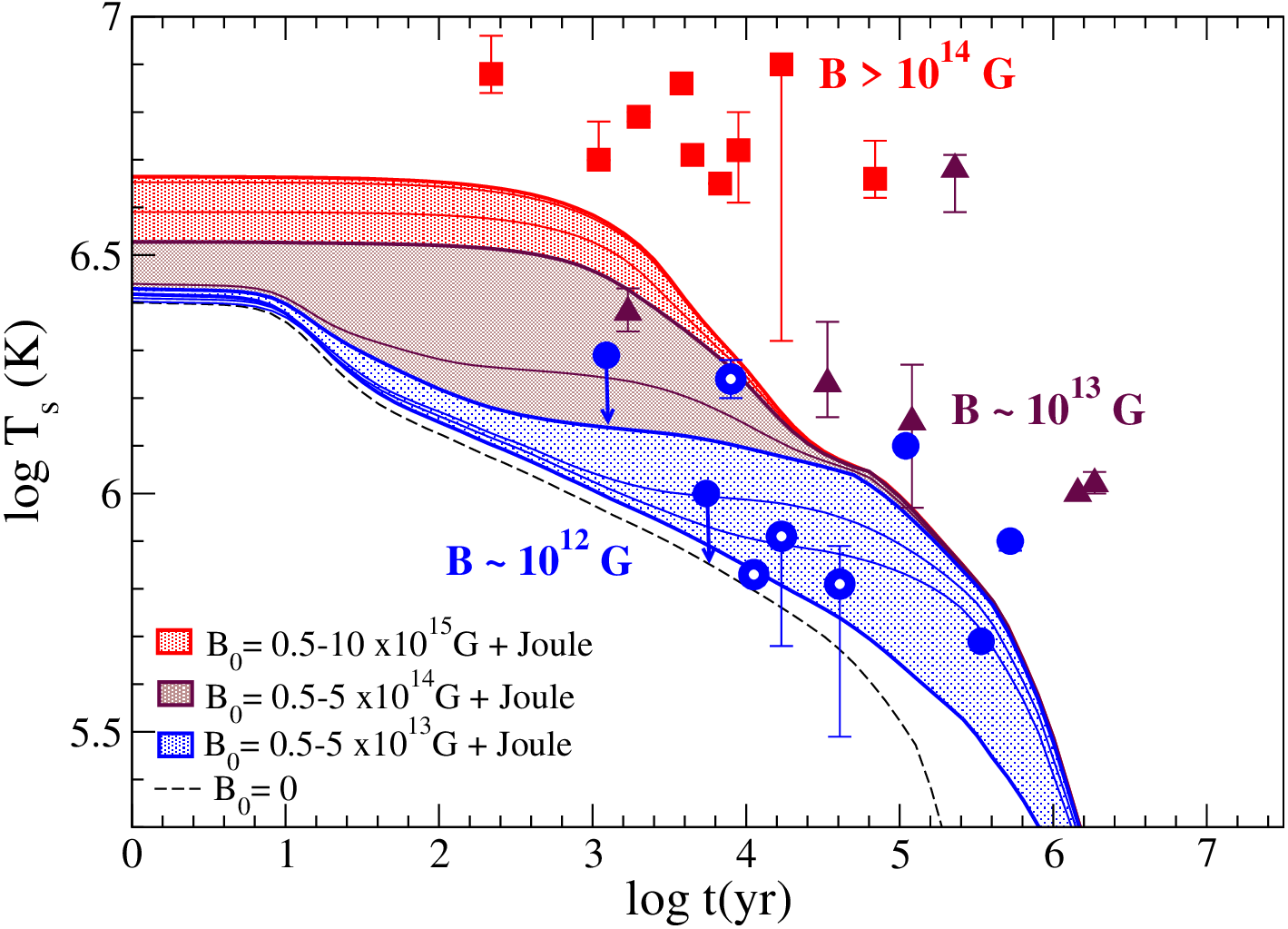}\\
\vspace{0.95cm}
\includegraphics[width=0.45\textwidth]{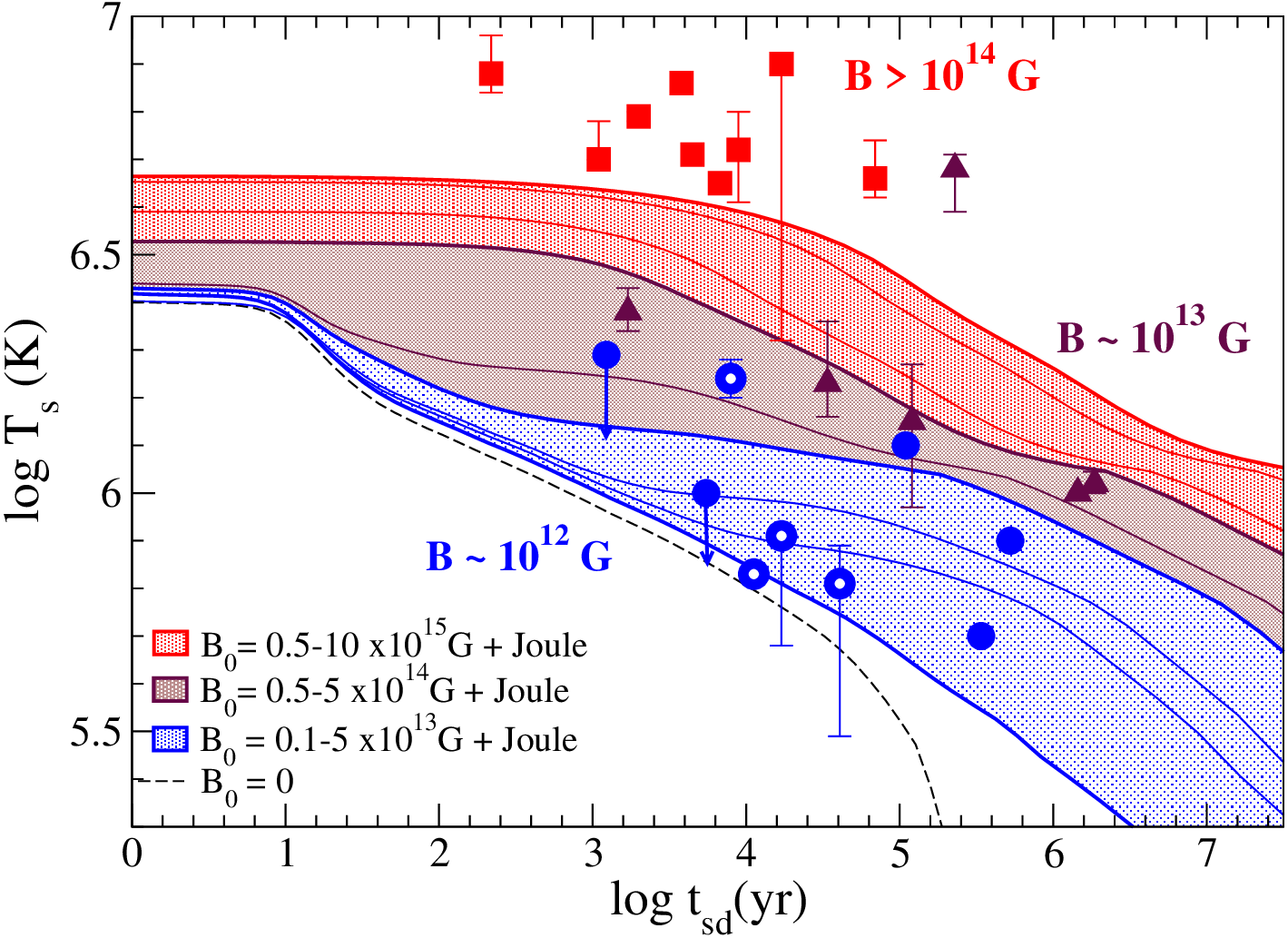}
\caption{Cooling curves. Surface temperature at the pole as a function of $t$
(upper panel) and $t_{sd}$ (lower panel) for different $B_0$.
The magnetar region (red) contains curves for
NSs born with $B_0 > 5 \times 10^{14}$G. 
Intermediate field NSs ($5\times 10^{13}$G$\,<B_0< 5\times10^{14}$G) 
evolve within the brown area and the blue region corresponds to NSs with 
$B_0 < 5 \times 10^{13}$G. 
Observations: squares for sources with $B_{d}>10^{14}$G, 
triangles with $10^{13}$G$\,<B_{d}<10^{14}$G, 
and circles with $B_{d}<10^{13}$G. Open circles are temperatures obtained from
fits to Hydrogen atmospheres.}
\label{Joule}
\end{figure}

In Fig.~\ref{Joule}, we show two types of theoretical curves: temperature versus
 true age (according to the simulation) and temperature versus the spin-down age consistent
with the assumptions made on each model about the magnetic field evolution.
We have sampled the objects in three groups
according to their measured magnetic field: {\it high field} NSs 
with $B_{d}\ge10^{14}$~G, 
{\it intermediate field} NSs with $10^{13}$~G$<B_{d}<10^{14}$~G and 
{\it low field} NSs for which $B_{d}\le10^{13}$~G.  
We found that this three samples could be explained qualitatively by cooling curves 
in three different regimes: high, intermediate and low magnetized NSs, in all cases with 
Joule heating included. Each of these regimes is represented by a set of curves with a 
given order of magnitude of the initial field, $B_0$. We have taken 
${\tau_{\rm Ohm}}=10^6$ yr and ${\tau_{\rm Hall}}=10^3 {\rm yr}/B_{0,15} $,
where $B_{0,15}$ is $B_0$ in units of $10^{15}$ G. The dependence of the results on the decay rates is 
discussed in \cite{Aguilera2007}.
These three regions are depicted in Fig.~\ref{Joule} as colored zones. For comparison, 
the cooling curve corresponding to a non-magnetized NS is represented by a dashed
line. 

Focusing in the high field (red) region, we see that the effect of the magnetic field is visible
from the very beginning of the NS evolution. The initial equilibrium temperature reached
in a non--magnetized model may be increased up to a factor of 5 and it is kept nearly
constant for a much longer time, up to $10^4$ years.
The effect of Joule heating is very significant and may help to understand 
the high temperatures observed in magnetars \citep{Kaspi2007},
 although other physical processes could contribute as well.
For instance, the initially higher temperatures result in higher electrical 
resistivity, therefore accelerating
the magnetic field dissipation in an early epoch. We have not yet considered the
consistent temperature dependence of the resistivity, in combination with the 
evolution of the magnetic field geometry. A fully coupled magneto-thermal evolution
code should be employed for this purpose (work in progress).
The observed temperatures of  radio-quiet, isolated NSs 
could be explained  either if they are old NS ($\approx 10^6$ years) born as magnetars
with $B_0=10^{14}$-$10^{15}$~G, or by middle age NSs born with fields in the range of 
$B_0=10^{13}$-$10^{14}$~G, as plotted in the intermediate brown zone of Fig.~\ref{Joule}. 
For intermediate field strengths, the initial effect is not so pronounced but the star
can be kept much hotter than non-magnetized NSs during the period from $10^4$ to $10^6$~yrs.
Finally, the effect of the magnetic field in stars born with initial fields of the order of
$10^{13}$ G is moderate, but still visible. These would be the case of some radio-pulsars, 
for which the detection of a thermal component in their spectrum has been reported.
For weakly magnetized NSs with $B \simeq 10^{12}$~G, the effect of the magnetic
field is small and they can satisfactorily be explained by non-magnetized models,
except for very old NSs ($t> 10^6$ years), as discussed in \cite{Miralles98}.
\begin{figure}[thb]
\vspace{0.8cm}
\centering
\includegraphics[width=0.45\textwidth]{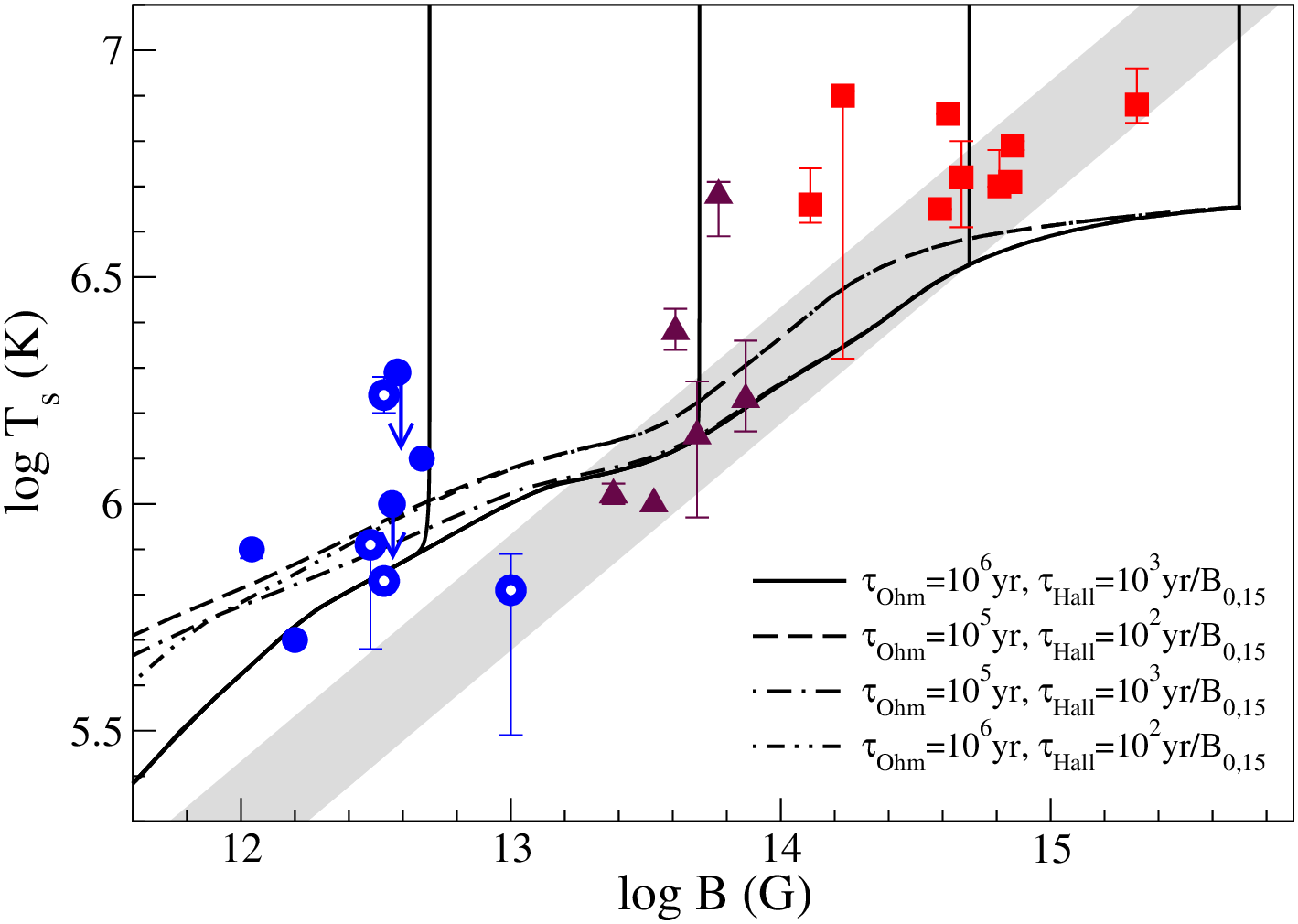}
\caption{Surface temperature (at the pole) as a function of the  
magnetic field. Theoretical curves are shown for different $B_0$ 
and varying $\tau_{\rm Ohm}$ and $\tau_{\rm Hall}$.
The grey band corresponds to the HBL \citep{PonsLink2007}. Observations correspond
to Fig.~\ref{Joule}.}
\label{FigTB}
\end{figure}

The fact that the magnetic field plays a relevant role in the thermal evolution of neutron
stars at least during the first million years of its life is supported but another observational
fact. The magnetic field distribution observed in radio-pulsars is peaked in $3\times 10^{12}$
G. Therefore, assuming no correlation between $B$ and $T$, one would expect to find 
thermal emission from neutron stars below and above
that value in approximately the same number.
However, almost all nearby objects for which its temperature is known have fields
far exceeding that value. This mismatch between the pulsar field distribution and the
thermal emitters field distribution can be easily explained by our assumption: magnetic
field decays significantly during the first $10^6$ years of a NS life and this effect governs its
thermal evolution. As a consequence, NSs with larger fields have higher temperatures
and cool slower,
increasing the chances to be detected. This hypothesis has been recently outlined by
\cite{PonsLink2007} who reported a strong correlation between
the surface temperature and the magnetic field, well approximated by the expression
\begin{equation}
T_{s,6}^4 \simeq C B_{d,14}^2
\label{eqHBL}
\end{equation}
where $T_{s,6}$ is $T_s$ in units of $10^6$ K, $B_{d,14}$ is $B_d$ 
in units of $10^{14}$ G, and $C$ is a constant that depends on the 
thickness of the crust, the Ohmic dissipation
timescale, and the ratio between the unknown internal field and the observed external
dipole. For typical numbers, $C \simeq 10$. This straight line in a logarithmic
plot of temperature versus magnetic field has been named the {\it heat balance line}
(HBL).

To contrast our results with this hypothesis based on a simple energetics argument, 
we plot in Fig.~\ref{FigTB} the variation of the polar surface temperature $T_s$ 
(usually identified with the hot component of the thermal spectrum)
as a function of $B$ for different theoretical curves, and we compare
them with the observational data from Table 1.
In this $T$-$B$ diagram the thermal history of a NS proceeds as follows.
A NS begins its life high on the figure with some initial field $B_0$. As it cools it moves
vertically downward, until decay of its field provides an energy source able to
counterbalance the thermal loses. This causes the trajectory to
bend to the left. Then it 
continues moving down but the temperature and the field evolution are coupled
since heating produced by magnetic field depends on the field strength and the
cooling mechanisms depend on the temperature.
For young magnetars the initial stage is very short because the high field prevents the star to cool down further and  the surface temperature remains relatively high. A further evolution proceeds only when the field is also decaying. For low magnetic field stars the drop in the initial temperature is more 
pronounced and they reach the equilibrium region much later. 
The region in the diagram where all cooling curves tend to converge is in agreement with 
the HBL defined in \cite{PonsLink2007} (indeed it is a band, since not
all NSs are necessarily identical).
The grey band corresponds to the choice $C=5$-$50$.

Since the initial vertical trajectory is very fast, NSs do not spend much 
time in that region of the diagram.
A clustering of sufficiently old objects in the grey region is expected.
Low magnetic field stars reach undetectable low temperatures more quickly, thus they
are more difficult to detect than highly magnetized NSs. 
In the late photon cooling era, since the main mechanism to radiate energy is
photons from the star surface, one expects that the correlation
given by Eq. (\ref{eqHBL}) is more clear. 
Although it was a very simplified approximation, we show that our evolution curves 
approach the HBL slope asymptotically. 
Finally, it is worth to notice that some objects are crossed by curves 
belonging to different initial $B_0$, so there is no unique 
determination of the initial conditions.

As a result of this investigation we conclude that consistent magneto-thermal evolution 
simulations are needed  before we can disentangle the properties of the interiors 
of neutron stars by studying their cooling history. A first step in that direction 
has been given in this work, showing that the effect
of magnetic field decay in the highly resistive crust (as opposed to the highly 
conductive core)
could be very large. It certainly has a significant impact  on the thermal evolution of stars
with $B \gtrsim 10^{13}$ G.
The thermal and magnetic evolution of neutron stars is (at least) a
two parameter space, in which the evolutionary tracks of NSs born with 
different initial conditions (the mass of the object, the initial magnetic
field) cannot be properly described by a unique temperature versus age 
curve. It becomes clear that the {\it minimal model} that hopefully will be used
to understand neutron stars needs to include the magnetic field structure and evolution.

\acknowledgements
D.N.A is supported by the VESF fellowship EGO-DIR-112/2005.
This work has been supported in part by the Spanish MEC grant AYA 2004-08067-C03-02.


\end{document}